# TIFR Zero-Pressure balloon program crosses a milestone

D. Anand[1]*, B. Suneel Kumar[1], Devendra Ojha[1,2]

1. TIFR Balloon Facility, ECIL Post, Hyderabad-500062, India.

2. Tata Institute of Fundamental Research, Colaba, Mumbai-400005, India.

* Corresponding author, D. Anand, email: anandd@tifr.res.in

 Co-author, B. Suneel Kumar, email: bsuneel@tifr.res.in

 Co-author, Devendra Ojha, email:  ojha@tifr.res.in

**Abstract**

High altitude scientific balloons offer unique opportunities to carry scientific payloads to stratospheric altitudes at a cost several orders of magnitude lower than corresponding satellite missions. Balloon-borne payloads are easy to implement allowing quick experiment turn-around times and inexpensive reflights can be conducted as payload is most of the times recovered. In addition, in-situ and high resolution spatial and temporal measurements of the earth's atmosphere can be made that might not be feasible with satellites. It is also used as a testbed to prove technologies for future satellite missions. Scientific ballooning was initiated at the Tata Institute of Fundamental Research (TIFR) in the year 1945 when scientific instruments were flown to stratospheric altitudes using a cluster of weather balloons for cosmic ray research. The need to have balloons float at constant stratospheric altitudes for astronomy studies led to the initiation of work on design and fabrication of Zero-Pressure polyethylene (ZP) balloons at TIFR in 1956. Since then several ZP balloon flights were conducted for studies in astronomy, atmospheric science, astrobiology, balloon technology and space technology development, leading to several important scientific results. In the year 2018, TIFR balloon program crossed an important milestone of conducting more than 500 ZP balloon flights. This paper presents recent advancements made in some areas of scientific ballooning and the details of balloon experiments conducted in the past two decades.

Keywords: zero-pressure balloon, stratospheric altitudes, cold brittle point, gore.

**Introduction**

The Indian ZP balloon program started by TIFR established a permanent Balloon Facility (BF) at Hyderabad (latitude $17.5^0$ N, longitude $78.5^0$ E, geomagnetic latitude $8^0$N) India, in the year 1969, with an aim to design, manufacture and launch scientific balloons upto stratospheric altitudes for research in various science disciplines. The scientific advantage of this location is the high cosmic ray vertical cut-off rigidity of 15 GV and the easy accessibility to both the northern and southern skies from balloon altitudes making it ideal for Astronomical studies. Several countries worldwide like the USA, France, Japan, Australia, Sweden, Italy, Brazil, etc., have balloon programs and facilities for scientific ballooning. However, BF is unique in the sense that, it carries out balloon design and manufacture, gondola fabrication and payload integration, provides Telecommand, Telemetry, Tracking (TTC) support and balloon launch - all under one roof, in addition to payload recovery. In other words it provides an end-to-end solution for scientific ballooning. Balloon experiments are conducted not only for TIFR, but also for several national laboratories and universities from India and abroad as well. BF has the capability to design and manufacture a range of balloon sizes to accommodate various payload mass for desired float altitude. The ZP balloon flights are conducted on a regional basis due to the restrictions in allocated air corridor by the civil aviation authorities, for balloon flight operations. Therefore, a float duration upto ~10 hours can be achieved based on wind speed and direction at the float altitude. Additionally, a maximum of about 400 km line-of-sight reliable range of the TTC system puts a limitation on the float duration. In the following sections, we will briefly dwell on balloon technology, support instrumentation, facilities available to scientists, our future plans and focus more on the balloon experiments conducted in the past two decades.

**Developments in balloon technology**

To manufacture a reliable ZP balloon for stratospheric altitudes, it is important to have a good balloon film, load tapes as per the balloon design, balloon adhesive tape, balloon end fittings and quality control during balloon manufacture process. The balloon manufacture at BF is completely indigenous in all these components. An exhaustive detail on our early balloons, balloon design and optimized single cap balloon design is discussed elsewhere[1-4]. Developments made in some of these areas are discussed.

*Indigenous balloon film*

In the mid-eighties, both M/s Winzen and Raven of the USA introduced new grades of balloon films viz. SF-372 and Astrofilm-E2 respectively with Cold Brittle Point (CBP) around -96ºC and exceptional mechanical properties. Inspired by this development, BF extruded balloon grade film out of "LADENE 118W" Linear Low Density PolyEthylene (LLDPE) resin with butene as comonomer having CBP around -88ºC and the mechanical properties were lower than SF-372 and Astrofilm-E2. Balloons using this film were successfully flight-tested in 1989. In 1990, superior grade LLDPE resin "Dowlex 2045" with octene as comonomer became available in India and the film extruded out of this resin was named Antrix[1]. Laboratory tests of Antrix film revealed its exceptional mechanical properties both at room temperature and at -80ºC, and the CBP of this film was found to be about $-96^{0}C^{2}$. Further improvements were made in manufacturing process during extrusion leading to better quality of Antrix film. The film is comparable with the best balloon grade films from USA like SF-372 and Astrofilm-E2. The arrival of Antrix film has been a game changer for scientific ballooning in India, resulting in 100% success for balloons flown from BF. Currently, all balloons fabricated at BF are made out of Antrix film. Details on the test methods and tools to determine the indigenous Antrix film's mechanical properties and comparison with SF-372 and Astrofilm-E2 is discussed elsewhere[4].

*Indigenous load tapes*

During balloon fabrication, balloon film is cut into many banana-peel shaped vertical sections called gores. The gore edges are heat sealed together along with the load tape made of high tenacity polyester yarn. Many such gores heat sealed together form a balloon envelope. During the balloon flight, the longitudinal stresses due to suspended load are carried by the load tapes. Therefore, load tape is the second major and expensive component in the balloon manufacture which was imported in the initial years. A load tape manufacturing unit indigenously developed after years of research and development is fully operational and by the year 2000 we were capable of manufacturing load tapes of Breaking Strength (BS) upto 136 kgf. However, load tapes for heavier Infrared astronomy telescope was imported. After a compelling search, a polyester yarn of 9.3 g/denier was identified and a load tape made out of this yarn, at BF, resulted in a BS of about 182 kgf. Balloons made of this load tape proved to be safe to carry the Infrared astronomy telescope making us totally self-reliant[5]. Currently, BF is capable of manufacturing load tape of breaking strengths upto 400 kgf.

*Indigenous balloon adhesive tapes*

Another critical component is the adhesive tape. This is used for reinforcement of the balloon top end fitting, hydrogen inflation tubes and for carrying out balloon repairs. The adhesive tape that was earlier imported is now manufactured in India. In the initial stages of development the manufacturer worked closely with balloon group to customize it for this application. The indigenous tape made out of Fluorinated Ethylene

Polypropylene film of thickness 50 µm coated with a silicone based adhesive, has all the properties required for balloon applications[5].

**Balloon support instrumentation capabilities**

The balloon flights are supported by high reliability telecommand uplink for balloon and payload control operations including flight termination, and telemetry downlink for payload health and science data. The S-band 3.7 m Dish TTC ground station provides precise autotracking of the balloon and is capable of handling telemetry data rates upto 1Mbps in the standard Inter-Range Instrumentation Group format used in ballooning. A fair amount of redundancy is provided in important subsystems to make flight operations foolproof. High reliability air safety devices are used to meet compliance with civil aviation rules. Payload is recovered within 1 to 2 hours of landing due to reliable tracking devices. In the past, several custom-designed and experiment-specific onboard electronics were developed to cater to the needs of user scientists. Gondolas of various sizes and shapes were designed and fabricated at BF to provide structural stability to payload during launch as well as landing.

**Science and Technology- development balloon flights (1999-2018)**

The first ZP balloon flight in India was conducted in 1958[3]. Up till 2018, 507 ZP balloon flights have been conducted for research in various science disciplines. Balloon flights with gross lifts upto 2500 kg with payload weight upto ~1000 kg having dimensions upto 3 m diameter and 5.5 m height can be carried out. Currently, the largest balloon that can be manufactured at BF is 740,000 $m^3$. Payload recovery is almost 100%. 71 successful balloon experiments conducted in various disciplines during this period are briefly discussed.

*X-ray Astronomy*

A telescope named Large Area Scintillation counter Experiment (LASE) was designed to detect microsecond variations in the flux of X-ray sources in the hard X-ray energy range up to 200 keV. Nine balloon flights were conducted during 1999-2012 using this telescope at float altitudes of 41-42 km. A large number of papers with important results on the rapid variability of galactic and extragalactic X-ray sources in hard X-rays have emerged from these flights[6,7]. Later the same group developed a new balloon-borne High Energy X-ray Imaging Telescope (HEXIT) for hard X-ray studies in the energy range of 20-800 keV which was successfully flown in 2005, 2006 and 2009. The second HEXIT payload, shown in Fig. 1, was launched aloft a balloon of volume 738,900 $m^3$ in the year 2006. The balloon floated at an altitude of about 41.4 km for about 1 hour and 21 minutes, with a suspended load of 938.2 kg[8]. The third HEXIT flight launched on April 09, 2009 using the same balloon size floated at an altitude of 41.7 km for about 9 hours. This is the second largest balloon designed, fabricated and launched from BF. Another type of X-ray balloon experiment comprising two Large Area X-ray Proportional Counters (LAXPC), was carried out on April 14, 2008. The design of these LAXPCs was similar to those on the ASTROSAT satellite, except that their field of view was $3^0 \times 3^0$ as compared to the $1^0 \times 1^0$ for the LAXPCs on the ASTROSAT. This experiment was aimed at timing and spectral studies of X-ray sources in 3-80 keV region. The black hole X-ray binary Cygnus X-1 was observed in the experiment for ~ 3 hours at a float altitude of about 41 km[9]. This payload was carried aloft a 739,500 $m^3$ balloon designed and fabricated at BF and is the largest balloon launched from BF till date, with a suspended load of 954 kg. The X-ray experiments described were conducted by the TIFR X-ray astronomy group.

*Infrared Astronomy*

As a part of the TIFR-Japan collaboration in balloon-borne Far Infra-Red (FIR) astronomy, a Fabry-Perot Spectrometer (FPS) developed by the Institute of Space and Astronautical Science, Japan, was successfully installed at the focal plane of the TIFR 100-cm balloon borne telescope. This new Telescope combination shown in Fig. 2 will now be referred as FPS100. The FPS (R~1800) is tuned to the astrophysically interesting fine structure line of [C II] at 157.74 µm. [C II] 158 µm line is a major gas cooling channel of the neutral gas-phase Inter-Stellar medium (ISM) and thus is a key to understanding ISM properties influenced by the radiative feedback of massive stars. The first successful flight of FPS100 was conducted on November 25, 1999[10]. Till date the FPS100 weighing ~1080 kg with ballast has been successfully flown 10 times since its first launch. Thanks to recent continuous successful observations, unprecedented large-area (~15 arcmin x 30 arcmin) [C II] maps of Galactic carbon line was made from several Galactic massive star-forming regions (viz. RCW 38, Orion Bar, W3, RCW 36, NGC 2024, W31, Carina, NGC 6334 and NGC 6357)[11]. Under another TIFR-Japan collaboration, five successful flights were conducted from 2000 to 2003 using the Japanese Far Infra-Red Balloon Experiment (FIRBE) telescope. FIRBE developed mainly by the Japanese Nagoya University team and partly by TIFR Infrared astronomy group has a 50 cm off-axis paraboloidal mirror at the focus of which a 32 element mechanically stressed photoconductor detector array at 155 µm is mounted. Interesting results from the wide-area mapping of the 155 µm continuum emission in the Orion molecular cloud complex were obtained[12].

*Atmospheric Science and Astrobiology*

The period under discussion, has seen a revival of balloon-borne experiments for atmospheric science research. Many of them are first of its kind experiments from BF.

A low power, high pulse repetition frequency laser based Micro Pulse Lidar operating at 532 nm having 150 mm Cassegrain type telescope and a photomultiplier tube based photon counting data acquisition system was developed by ISRO for Balloon Borne Lidar (BBL) based measurement of backscattered signal from aerosols and clouds. This downward looking 459 kg BBL was successfully launched using a 109,700 $m^3$ balloon on April 16, 2009 at 00:36 IST[13]. The flight was terminated after a float duration of 2 hours at 35 km altitude. This is the first time that a BBL was flown upto stratospheric altitude, from BF.

An Indo-US Balloon-borne Investigation of Regional-atmospheric Dynamics (BIRD) experiment was conducted on March 08, 2010. The 109,755 $m^3$ balloon carrying a 335 kg BIRD payload reached a float altitude of about 35 km at 12:45 IST. The flight was terminated at 18:25 IST. The Gondola orientation, designed by BF, was programmed to rotate in azimuth direction in a unique fashion to achieve the experimental objectives. The pointing accuracy of the orientation platform attained was 20 arcminute. This is perhaps the first daytime balloon-borne optical investigation carried out on mesosphere-lower troposphere wave dynamics[14]. This is also the first time that a balloon was launched around noon from BF.

An Aethalometer (Model AE-42, of Magee Scientific, USA) was flown aloft a balloon of volume 109,755 $m^3$ on March 17, 2010. The payload weighing 350 kg reached a float altitude of 35 km. The balloon ascent rate for this flight was maintained at ~2.6 m/s to enable high resolution vertical profiles. This is the first ever in-situ measurements of black carbon aerosols in the troposphere (up to 9 km) made over central India and was conducted by the Space Physics Laboratory, Thiruvananthapuram, India[15]. The same experiment was repeated twice later in 2011 to continue the investigations further.

A Balloon Experiment on the Electrodynamics of Near Space (BEENS) was launched using an 110,000 $m^3$ balloon on December 14, 2013. The balloon reached float altitude of 35.2 km and the flight was terminated after 4 hours of float duration. The payload comprising four deployable booms for measurements of horizontal electric fields and one inclined boom for vertical electric field measurement, were successfully deployed through telecommand, after launch[16]. A unique orientation mechanism for the experiment was designed by BF engineers. The experiment was conducted by the Indian Institute of Geomagnetism, Navi Mumbai, India.

The Balloon measurement campaigns of the Asian Tropopause Aerosol Layer (BATAL), is a NASA–ISRO sponsored campaign in India and elsewhere to study the nature, formation, and transport of polluted aerosols in the upper troposphere and lower stratosphere during the Asian summer monsoon season[17]. BATAL ZP balloon flights were conducted in 2015 and 2017-2019 from BF. This is the first time that the balloons were designed and fabricated by BF for the upper troposphere and lower stratosphere float altitudes and medium payload mass. The onboard telemetry, telecommand, GPS interface, air safety transponder and batteries were custom designed by BF to keep the weight as low as possible (11 kg). A miniaturized ballast system was designed that helped achieve the balloon to float about the tropopause region. This was also the first time that ZP balloon launches were conducted during the Indian summer monsoon season when high surface winds prevail, making launch very challenging. A special permission was sought from the airport authorities due to the proximity of the balloon float altitude to aircraft cruise altitude (~12 km).

Three balloon borne experiments were conducted in the area of astrobiology that involved many astronomers and biologists. The cryogenic sampler payload was developed by ISRO. The first one was launched on April 29, 1999, the second one on January 21, 2001 and the third one on April 20, 2005. Air samples were collected aseptically at altitudes of 20-41 km using a cryogenic sampler comprising a 16-cryoprobe assembly, for studies on the qualitative and quantitative distribution of micro-organisms in the upper troposphere-stratosphere. Few new and novel bacterial strains have been found in the samples collected [18,19].

*Balloons for mesospheric altitudes*

The work on development of ultra-thin balloon grade polyethylene film, using m-LLDPE resin with added metallocene catalyst, in thickness range 2.8 μm to 3.8 μm for fabricating high altitude balloons capable of penetrating the mesosphere commenced in 2011 to meet the needs of scientists working in the area of atmospheric dynamics. After initial setback in 2012, three flights were conducted in 2014, using 61,013 m³ balloons made out of 3.8 μm film with a suspended load of about 10 kg. All the three balloons crossed into mesosphere reaching altitudes of over 51 km and the highest altitude reached is 51.83 km[20].

*Balloon for StratEx human Spaceflight record*

On October 24, 2014, Dr. Alan Eustace ascended to an altitude of 41.578 km under a 328,232 m³ balloon – the greatest height a person has ever reached without rocket propulsion. As the balloon descended to an altitude of 41.422 km, Dr. Eustace detached himself from the balloon with the aid of a pyrocutter device and plummeted toward the earth at speeds that peaked at 821 miles per hour, setting off a  new world altitude record for a skydive, which stands unbroken till date. The balloon for this flight was designed and manufactured at BF using indigenous Antrix film and was launched by the StratEx team from New Mexico, USA. A quote from the paper "*The team feel that TIFR's proprietary films and balloon assembly are of the highest standard and had no qualms about the safety of launching Alan Eustace under these balloons*"[21], speaks for itself on the quality of our balloons.

*Animal Spaceflight*

A capsule with three live lab rats with life support system for its survival was launched aloft a 39,915 m$^3$ balloon on March 14, 2015. The flight was terminated after the balloon reached an altitude of 29.5 km. The design of life support system was successful as all the rats were recovered alive and healthy. BF engineers contributed in making the capsule leak-proof. The experiment was carried out by InGenius, Singapore — the company behind the project. This is the first time such an experiment was conducted from BF.

*Engineering test flight – 500$^{th}$ flight (T500)*

The FPS100 telescope which is in the process of being upgraded requires enhanced telemetry data handling capability. After considering several options, it was felt that the easiest way is to split the data and transmit them through two independent Radio Frequency (RF) transmissions. Currently, all telemetry data comes through a single RF transmission. The onboard electronics was designed to transmit two telemetry data on two different frequencies, one at the standard 10 kbs data rate on 2259 MHz and the other at 250 kbps on 2281 MHz. In order to flight-test the dual telemetry scheme and several other electronics, an Engineering test flight was conducted on April 13, 2018. The balloon of volume 3026 m$^3$ with a suspended load of 72 kg reached a float altitude of 24.9 km. The flight was terminated after float duration of 3 hours 20 minutes. T500 balloon before lift-off is shown in Fig. 3. A BF customized 6-stage Quartz Crystal Microbalance (QCM of California Measurements USA) impactor, which was first flown on May 07, 2016[22], was again flown in this flight to make aerosol measurements in the free troposphere at different altitudes. The telemetry was received by a common dish antenna, followed by two independent chain of data acquisition equipment in the control room. The performance of dual telemetry scheme, QCM including new onboard electronics were satisfactory. QCM data is being investigated. This is the first time that the dual telemetry scheme was successfully flight-tested and is likely to open up new opportunities for complex balloon experiments. *As a coincidence, on this date, the TIFR balloon program reached an important milestone of conducting 500 ZP balloon flights in India.*

The balloon flights conducted year-wise are shown in Fig. 4 and percentage distributions by various disciplines are shown in Fig. 5.

## Facilities

BF has a 1 m dia. x 1.5 m long Thermal-Vacuum chamber that operates in the temperature range of $-40^0$ C to $80^0$ C and a vacuum level of $3.3 \times 10^{-5}$ mb can be achieved and a programmable Hot chamber that can attain $+100^0$ C. For calibration of pointing system a non-magnetic suspension stand, two clean room facilities of class 10,000, four laminar air flow benches of class 100 and a Shaker for shock and vibration test are also available. There are 4 Labs for scientists to work on their payloads and a rigging area for Payload tests. A mechanical workshop, electronic test and measurement equipment are available for use to the scientists. Guest house and canteen facility is available within the campus.

## Future plans

As payloads are becoming more complex and heavier, design and development of ZP balloons that can carry payloads heavier than ~1000 kg to altitudes of ~38 km and above, has been initiated. Currently, ZP balloons designed by BF can carry ~1000 kg payload to an altitude of 32 km. Next, a ZP balloon is under development that can carry 5 ton payload to a float altitude of 14-15 km. These balloons find applications in testing the performance of re-entry capsules of human space-flight missions. In the recent times there has been an increasing demand for tethered Kytoons capable of hoisting 20-25 kg payload at 300 m altitude for testing communication payloads and to deploy internet services in remote areas for disaster management. To cater to these needs of national importance, design and fabrication of 100 $m^3$ Kytoon using a dual layer polyethylene film is under progress. We are pursuing development of Super-Pressure (SP) balloons as there is a growing need from the scientific community for long and ultra-long duration flights. A suitable balloon film has been developed and design and manufacture of SP balloon is in progress. The scientific ballooning community (from USA, Japan, France, Australia, etc.) has established necessary infrastructure at several launch bases spread across the world (Antarctica, Australia and Sweden- just to name a few) and have been conducting long duration flights, using ZP and SP balloons, ranging from few days to several weeks. Under collaboration with these countries and with substantial funding TIFR can implement such missions in future.

**Proposal for balloon experiments**

Proposal for balloon experiments can be sent to the scientist-in-charge over email. (https://www.tifr.res.in/~bf/ is our website). After examining the feasibility, further course of action will be communicated. A soft copy of Balloon Facility user's manual is available on request.

**Conclusion**

The importance of scientific ballooning for research in astronomy, atmospheric science, astrobiology and development of space technology is demonstrated beyond question. Several first of its kind balloon experiments were conducted from Balloon Facility. Payload recovery facilitated multiple reflights to be conducted leading to important scientific observations and results. Several national and international research institutes conducted balloon-borne experiments from BF. The X-ray detector which was tested on a balloon platform before being successfully deployed in AstroSat – a first dedicated Indian astronomy satellite mission, is providing excellent scientific results. In the past two decades, major innovations were made in all areas of scientific ballooning making BF self-reliant. New initiative to design and develop large Kytoons, super-pressure balloons for long duration flights and ZP balloons to carry heavier payloads will open up new opportunities to perform science and test space technologies.

**Acknowledgement**

The TIFR balloon program is greatly indebted to Dr. Homi J. Bhabha, Prof. M G K Menon, Prof. G S Gokhale, Prof. Bernard Peters, Prof. R R Daniel, Mr. R T Redkar, Prof. S V Damle and Mr. M N Joshi who pioneered the art and science of scientific ballooning in India. Our due thanks to Prof. P C Agrawal, Prof. R K Manchanda, Prof. S K Ghosh and Mr. S Srinivasan for their major contributions. It is a pleasure to thank our colleagues at Balloon Facility Hyderabad for their dedicated support. The authors acknowledge the support of the Department of Atomic Energy, Government of India, under Project Identification No. RTI 4002. We thank the anonymous reviewer for several useful comments and suggestions.

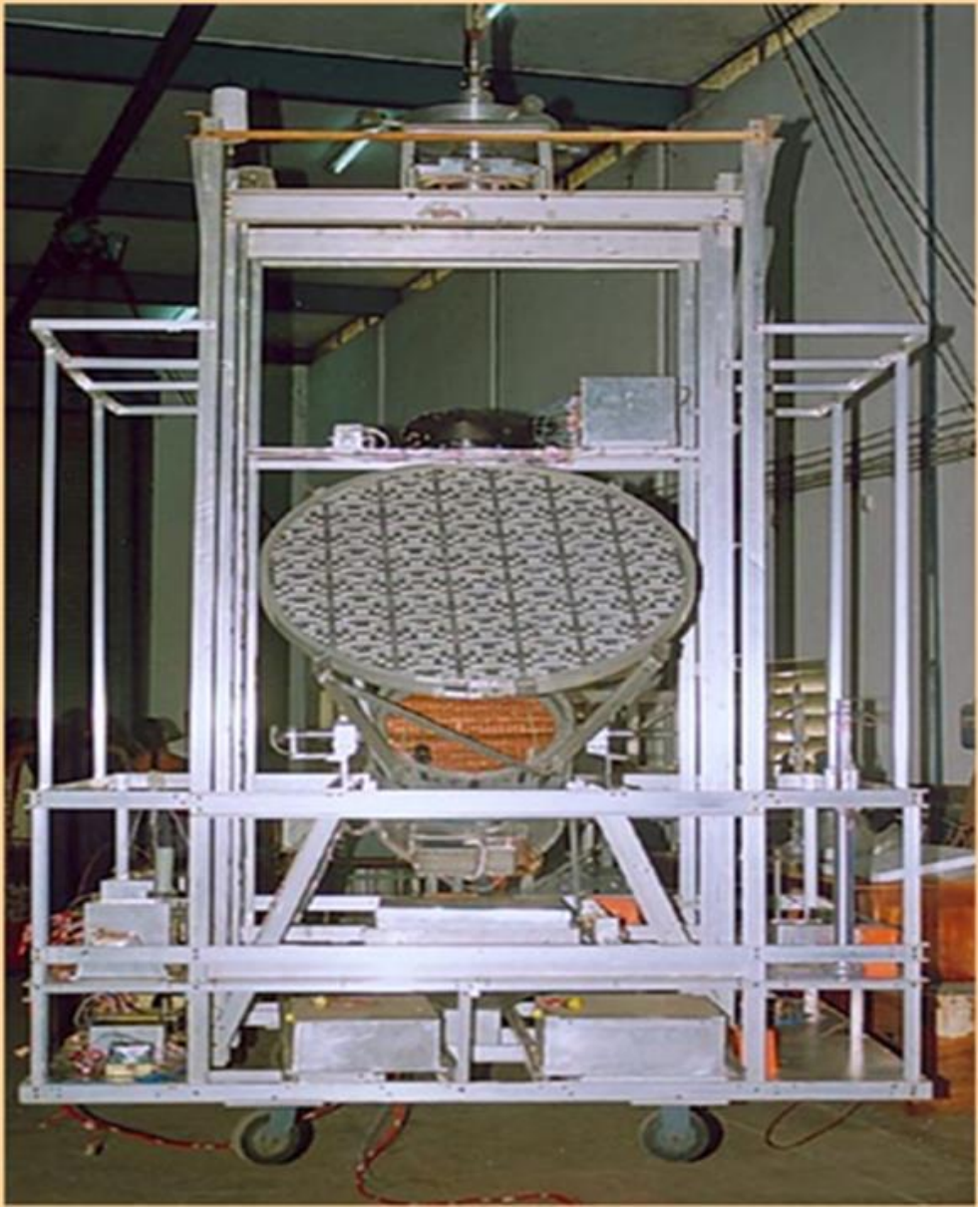

Figure 1.  X-ray Telescope – HEXIT.

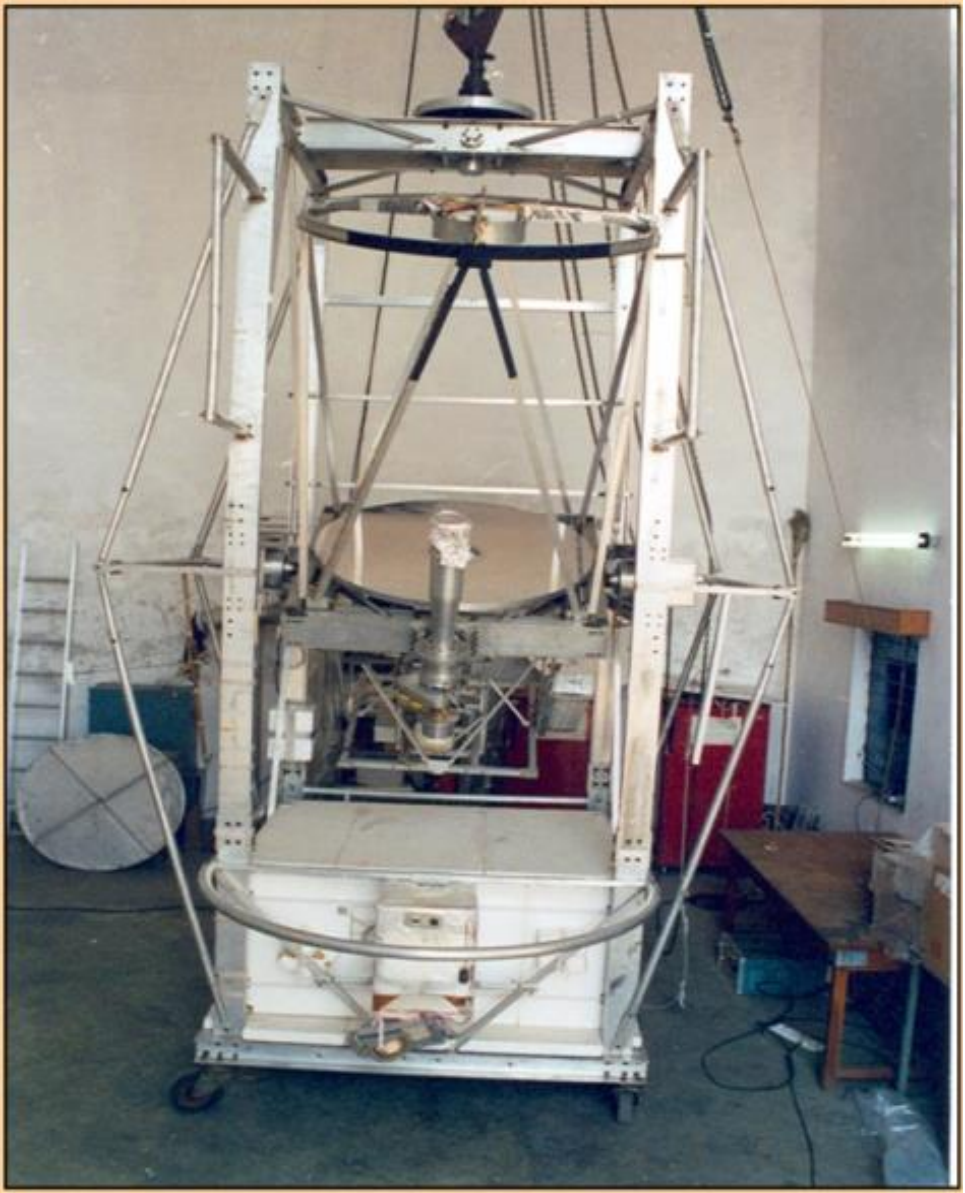

Figure 2. FPS100 - FIR Telescope.

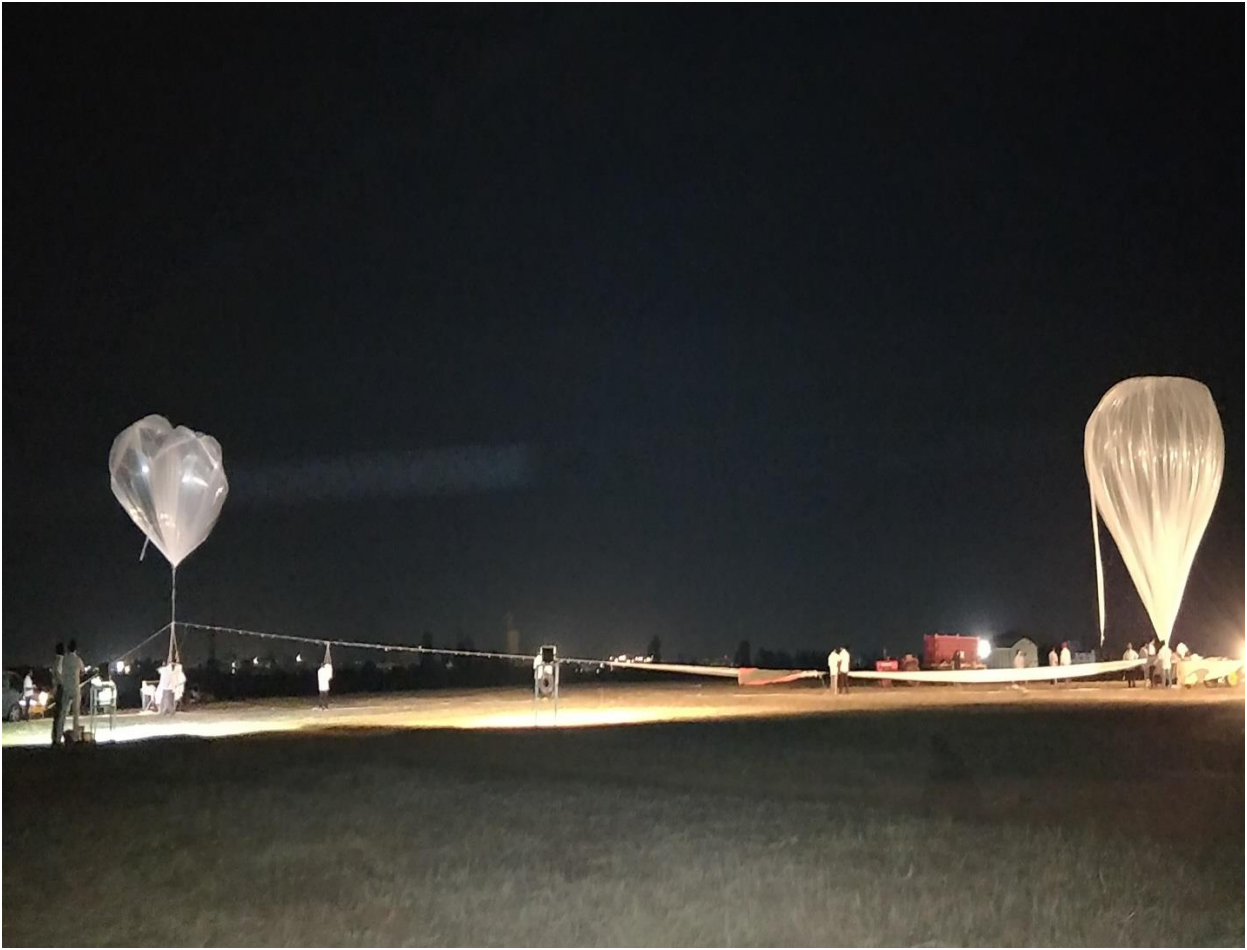

Figure  3. Hydrogen inflated T500 balloon before lift-off – static anchor line launch method.

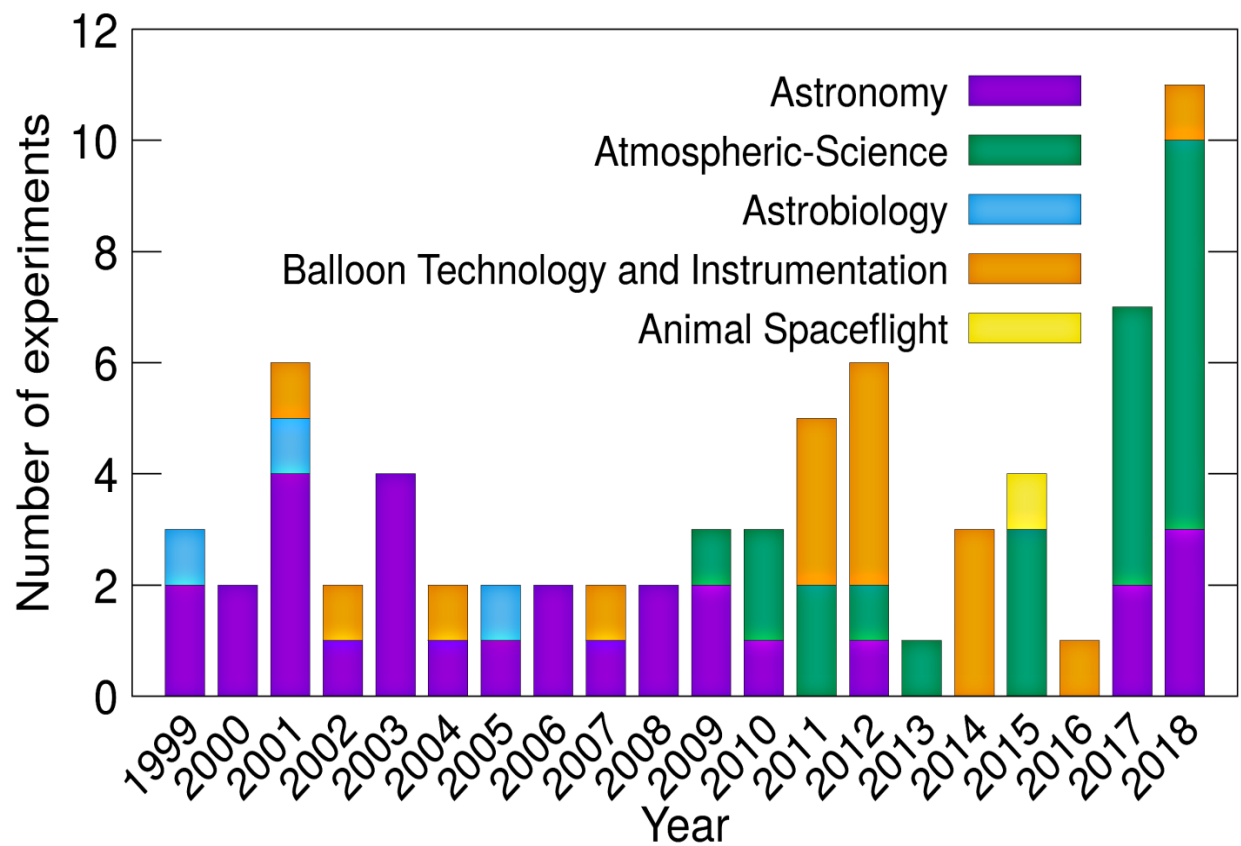

Figure  4. Balloon experiments conducted year-wise from 1999-2018.

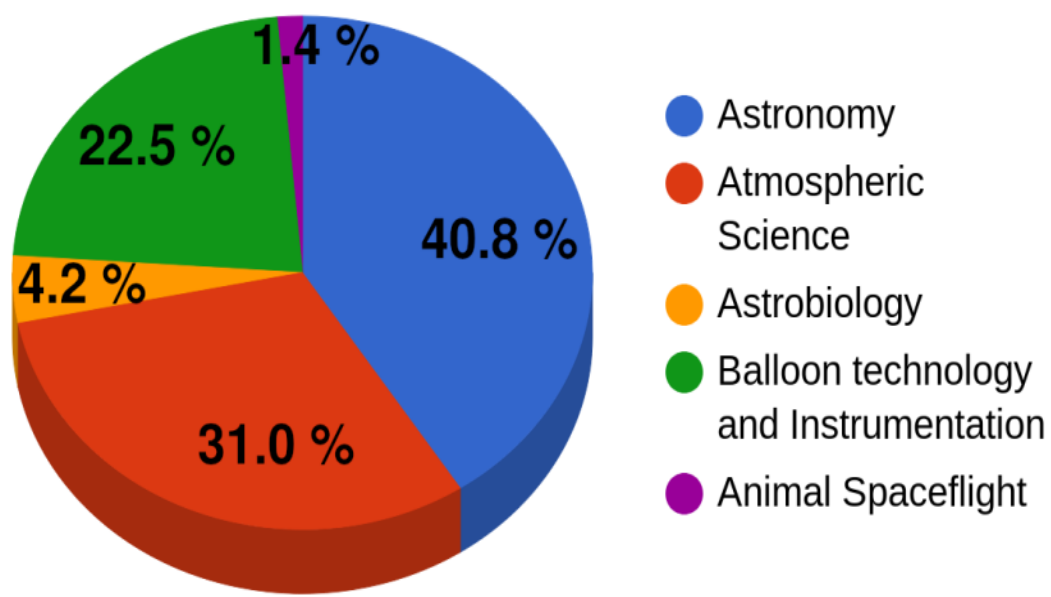

Figure  5. Percentage distribution of balloon experiments from 1999-2018.